\PassOptionsToPackage{table}{xcolor}
\documentclass[acmsmall]{acmart}

\AtBeginDocument{%
  \providecommand\BibTeX{{%
    \normalfont B\kern-0.5em{\scshape i\kern-0.25em b}\kern-0.8em\TeX}}}

\copyrightyear{2022} 
\acmYear{2022} 
\setcopyright{acmlicensed}
\acmConference[CHI '22]{CHI Conference on Human Factors in Computing Systems}{April 29-May 5, 2022}{New Orleans, LA, USA}
\acmBooktitle{CHI Conference on Human Factors in Computing Systems (CHI '22), April 29-May 5, 2022, New Orleans, LA, USA}
\acmPrice{15.00}
\acmDOI{10.1145/3491102.3517505}
\acmISBN{978-1-4503-9157-3/22/04}

\usepackage{tabularx}
\usepackage{color}
\definecolor{lightgray}{gray}{0.9}
\usepackage{array}
\usepackage{graphicx}
\usepackage{subcaption}
\usepackage{mwe}

\newcolumntype{L}[1]{>{\raggedright\let\newline\\\arraybackslash\hspace{0pt}}m{#1}}
\newcolumntype{C}[1]{>{\centering\let\newline\\\arraybackslash\hspace{0pt}}m{#1}}
\newcolumntype{R}[1]{>{\raggedleft\let\newline\\\arraybackslash\hspace{0pt}}m{#1}}

\begin{document}

\title{Designing Word Filter Tools for Creator-led Comment Moderation}






\author{Shagun Jhaver}
\affiliation{%
  \institution{Rutgers University}
  \country{USA}
  }
\email{shagun.jhaver@rutgers.edu}

\author{Quan Ze Chen}
\affiliation{%
  \institution{University of Washington}
  \country{USA}
}
\email{cqz@cs.washington.edu}

\author{Detlef Knauss}
\affiliation{%
  \institution{University of Washington}
  \country{USA}
}
\email{detlefk@uw.edu}

\author{Amy Zhang}
\affiliation{%
  \institution{University of Washington}
  \country{USA}
}
\email{axz@cs.uw.edu}

\renewcommand{\shortauthors}{Jhaver et al.}

\begin{abstract}
Online social platforms centered around content creators often allow comments on content, where creators moderate the comments they receive. As creators can face overwhelming numbers of comments, with some of them harassing or hateful, platforms typically provide tools such as word filters for creators to automate aspects of moderation. From needfinding interviews with 19 creators about how they use existing tools, we found that they struggled with writing good filters as well as organizing and revisiting their filters, due to the difficulty of determining what the filters actually catch. To address these issues, we present FilterBuddy, a system that supports creators in authoring new filters or building from existing filter lists, as well as organizing their filters and visualizing what comments are captured over time. We conducted an early-stage evaluation of FilterBuddy with YouTube creators, finding that participants see FilterBuddy not just as a moderation tool, but also a means to organize their comments to better understand their audiences.

\end{abstract}



\maketitle

\fbox{
\begin{minipage}{36em}
  \textbf{CONTENT WARNING}: This paper contains offensive language, including misogynistic slurs, that readers may find disturbing.
\end{minipage}
}

\section{Introduction}

\begin{quote}
    \textit{``Some of you have never had to schedule dedicated time each month in your calendar to remove all the `nice boobs'/`she's cute'/`can she actually code?'/`get back in the kitchen' comments from your technical videos on YouTube, and honestly? It shows. [Please] do not comment telling me to turn off comments...I like to get comments/nice things said to me (just like anyone else would)...And this is just YouTube. Don't forget Vimeo/my Instagram/Twitter/LinkedIn/everything else, too!} 

    -- Chloe Condon, 2021~\cite{condon}
\end{quote}

\noindent 
Platforms like YouTube, TikTok, and Twitch combine technical and social features to support the formation of a participatory culture in which ordinary individuals create, collaborate, and share personally meaningful media, often as a source of income~\cite{chau2010youtube,raj2021top,niebler2020youtubers}. 
With thousands of new creators joining these platforms every year, the competition among creators for attracting viewers is high \cite{holmbom2015youtuber}.
In addition to the content they produce, many creators distinguish themselves
by working to curate a vibrant and inviting community around their content through interacting with and moderating their audience's commentary.

However, as Chloe Condon, a content creator who makes technical tutorials on YouTube and elsewhere, addresses in her viral Twitter thread excerpted above, creators---in particular, ones from marginalized groups---must invest a great deal of emotional labor to address unwanted comments from their viewers~\cite{parkin2018youtube}.
As channels grow larger and attract more comments, it becomes increasingly difficult for creators to review every comment, and some automation becomes necessary.
Even for the small minority of creators who can hire staff or are supported by volunteer moderators, receiving abuse via comments can contribute to anxiety and post-traumatic stress disorder (PTSD) for creators due to the oftentimes personal nature of the attacks \cite{parkin2018youtube}.
In light of this, 
creators like Condon have frequently requested more powerful moderation tools and resources that can meet the needs of marginalized creators and alleviate the pressures surrounding content moderation.

In this paper, we investigate the design of moderation tools targeted towards content creators to understand how they can better address creators' needs, especially the needs of creators from marginalized groups who face disproportionate and targeted abuse~\cite{green2015lesbian}. We focus our needfinding and design exploration around \textit{word filter tools}, one of the most common moderation tools offered to creators on social media platforms, that allow creators to configure a list of blocked keywords.
Through needfinding interviews with 19 creators, we examine their experiences with receiving inappropriate comments and their daily experiences and strategies using existing word filter tools to try to reduce their moderation workload and stress.
We found that creators were overall frustrated with the rudimentary features provided by existing word filter tools on platforms such as YouTube.
Creators described having difficulties with both building up a set of useful filters from scratch, as well as organizing a growing list of filters and auditing what their filters were actually catching. 

To address these needs, we present FilterBuddy, a system built for YouTube creators that improves the usability of word filter tools for authoring, maintaining, and auditing word filters.
Users can hook the system up to their YouTube channel and use it instead of the native word filter tool on YouTube to moderate their comments.
Features of FilterBuddy include but are not limited to: interactive previews of what different filters would capture during the authoring process; the ability to build from existing filters created by others; organization of filters into spelling variants and higher-level categories; and time-series graphs and tables to understand what comments are caught by which filters over time.
FilterBuddy is not just a research tool; it is a system designed to be actively used by YouTube creators to enhance their moderation capabilities.
While this tool was designed to address the needs of marginalized creators and protect against online harm, we have found that it may also have a wider range of uses and benefits for a broader range of creators.


We conducted an exploratory qualitative user study of FilterBuddy with 8 creators on YouTube who interacted with the tool loaded with their YouTube comments while providing feedback. We found that participants were appreciative of greater automation but did not prefer to sacrifice their control over the tool's operations and requested additional defensive mechanisms to reduce incorrect removals.
They also considered the ability to share their filters and build off of filter lists created by others to be a powerful means to reduce toxic content across YouTube. In addition, we were surprised to find that participants devised many use cases for the tool that go beyond just moderating undesirable posts.
Overall, participants felt that the designs explored in FilterBuddy would empower content creators, especially those belonging to marginalized groups, to efficiently address their content moderation needs as well as better understand their audiences.

We conclude by discussing the tensions and tradeoffs in creators' goals that are important to consider when designing rule-based moderation tools. 
Our design exploration shows that creators are deeply invested in retaining control over their moderation operations; we reflect on how sensible design defaults can help creators achieve this goal while minimizing the manual effort of setting up granular filters.
We also highlight that trust-sensitive collective governance mechanisms will be required to resolve the tensions creators face between seeking to build on other creators' word filters while also preferring to keep their configurations private to avoid exploitation by bad actors. 
Finally, we had limited development resources to implement FilterBuddy---yet, we found that its features are seen as highly desirable by creators. In light of this, we call upon platforms to step up their efforts and investments in developing tools that improve creators' working conditions.
We also examine how policymakers, third-party developers, and minority support groups can offset the moderation workload of creators.



\section{Related Work}
We focus our review on content creators and the unique challenges they face, the nuances and difficulties of addressing online harm and safety, and the design of existing content moderation tools such as word filters.

\subsection{Experiences of Content Creators Online}
Rajendra-Nicolucci and Zuckerman define \textit{creator logic platforms} as platforms that ``enable users to share a specific type of media (like video, livestreams, or art), in a one-to-many fashion.'' \cite{raj2021top}.
Such platforms include YouTube, TikTok, and Twitch, which let people from around the world upload content online and accumulate relatively large followings \cite{abidin2015communicative,hatfield2018not}.
Dominant social networks like Facebook, Instagram, and Twitter are used for a broader variety of purposes, such as connecting with family and friends, but creators can use these networks to share their content with a large audience as well \cite{raj2021top}.

Content creators not only have to write, produce, and edit their content, but they must also cultivate and engage with their audiences \cite{wohn2020audience,hamilton2014streaming,hilvert2018social,kim2019will}.
Creators often adopt strategies of ``micro-celebrity,'' where they regard their audience as fans, perceive content delivery as a performative act \cite{li2019live,pellicone2017game}, 
and use strategic intimacy to appeal to viewers through direct interactions \cite{berryman2017guess,raun2018capitalizing,marwick2015you}.
The technical affordances of content sharing platforms, such as the ability for users to view, comment on, and share content, also encourage the development of active, networked groups \cite{abidin2015communicative,cunningham2017being,lange2007commenting}, where
content creators and viewers interact with each other \cite{burgess2018youtube}.

As a result, creators often find themselves overwhelmed with the varied demands of their work and can suffer occupational stress \cite{parkin2018youtube}. 
While a small minority of content creators become successful enough to hire production crews and content moderators, the vast majority do their best to manage different tasks by themselves \cite{wohn2020audience}.
Opaque algorithms used by these platforms evaluate content creators and shape their popularity and earnings, thereby rendering consistent earnings uncertain and risky \cite{pedersen2019my}.
Therefore, creators often hesitate to hire staff for help even though platforms encourage struggling creators to enlist support \cite{parkin2018youtube}.
Platform-enacted demonetization of creator posts that contain inappropriate comments are an additional concern for creators when conducting content moderation \cite{caplan2020tiered,burch_2019}. 
Many creators also experience doxxing, harassment, stalking, and online threats, which further contribute to their mental health risks \cite{lewis2021we,uttarapong2021harassment}.
Prior research shows that existing platform tools fail to address a range of creator needs, including community information needs~\cite{mallari2021understanding}.
In light of this, any tool that helps creators manage their channel can benefit not just the creators by reducing their workload and stress but also improve the experience of their audiences. 

\subsection{Online Harm and Safety in HCI}
Online harm and abuse are highly subjective, situated concepts \cite{scheuerman2018safe} whose interpretations vary across cultures \cite{jiang2021understanding} and even
across individuals \cite{jhaver2018view}.
Online platforms do not explicitly define these concepts in their policies and take an ad hoc approach to setting standards for addressing them \cite{pater2016,kaye2019speech}.
Scheuerman et al. identified four types of online harm---physical, emotional, relational, and financial \cite{scheuerman2021framework}.
Elsewhere, Scheuerman et al. showed that experiences of online harm can occur in varied ways, i.e., abuse aimed directly at the individual or witnessed by the individual; abuse perpetrated by those outside or within the individual's social circles; and abuse impacting a specific individual or larger communities \cite{scheuerman2018safe}. 
Content creators and their viewers can experience harm in all of these ways through comments posted on their channel.

The related concept of \textit{safety }is usually referred to as protection from emotional, physical and social harm that may or may not be caused by abusive behavior \cite{scheuerman2018safe,ringland2015making,herring2002searching,vitak2017,kenney2001mapping}. Prior HCI research on safety has largely focused on interpersonal harm, which includes hate speech, cyberbullying, and online harassment \cite{poland2016haters,campbell2005cyber,smith2006investigation,lwin2012stop,matias2015}, and how it causes emotional distress or threatens physical safety \cite{scheuerman2018safe}.
Redmiles et al. uncovered the multifaceted nature of online safety on Facebook, arguing that it involves not only digital privacy, digital security, and harassment, but also offline safety and a sense of community support as well as the upholding of community values \cite{redmiles2019just}.
Building upon this research, we examine the safety perceptions of content creators and
explore the development of a privacy-sensitive anti-harassment tool, containing features to foster community support, to increase the safety of content creators online.

One particularly distressing statistic related to online harm and safety is that
marginalized communities are disproportionately affected by internet-facilitated harassment and cyber-bullying \cite{jhaver2018blocklists,Mahar2018squadbox}. For example, in 2020, Blacks (54\%) and Hispanics (47\%) were more likely than Whites (17\%) to report victimization due to their race/ethnicity, and 47\% of women claimed to have been harassed online due to their gender as opposed to 18\% of men \cite{vogels_2021}. Prior research has also documented incidents of online abuse against members of the Black Lives Matter movement \cite{ince2017social}, women \cite{eckert2018fighting,nova2018silenced}, and LGBTQ communities \cite{fernandez2019don,scheuerman2018safe}. 
Such disproportionate abuse is also reflected in the experiences of content creators, e.g., female and LGBTQ Twitch streamers frequently encounter harassment \cite{uttarapong2021harassment,freeman2020streaming}.
Our needfinding interviews with creators from marginalized groups echoed these findings, and we built on this prior work to explore 
how creator-led moderation tools can empower creators from marginalized communities to resist interpersonal harm.

HCI researchers have also focused on content-based harm, which occurs as a result of viewing undesirable content on social media platforms. For example, this includes race-based trauma caused by viewing posts of police violence against Black people \cite{tynes2019race,williams2019desensitization,to2020they}.
Digital self-harm has been shown to involve users posting or seeking out self-harming content, such as pro-anorexic posts \cite{chancellor2016thyghgapp} or images of self-harmed bodies \cite{andalibi2017sensitive}.
Research on the work practices of content moderators has shown how continuous exposure to violent, hateful, or otherwise troubling posts takes an emotional toll, causes secondary trauma, panic attacks and other mental health issues, and eventually burns out many workers \cite{roberts2019behind,jhaver2019automated,wohn2019volunteer,dosono2019moderation,steiger2021psych}.
We describe in this work the kinds of content that creators want to filter out and present a system aimed at reducing the need to continually expose oneself and one's audience to content-based harm.


\subsection{Content Moderation Tools}
\textit{Content moderation} refers to a series of practices, including rule-setting, using human moderators, and deploying automated tools, that are used to screen user-generated content to decide what posts will appear on, or remain on, a platform \cite{gerrard2018,roberts2019behind,Carlson2020}.
Many platforms, including Reddit, YouTube, and Twitch, have multiple levels of content moderation---while the majority of the content moderation workload is outsourced to the content creators or volunteer community moderators who establish and enforce local rules, the company may also step in to override an individual creator or community's self-policing \cite{Carlson2020,jhaver2021designing}.
According to Section 230 of the Communication Decency Act (CDA), social media companies are not treated as publishers and, therefore, with some exceptions related to sex-trafficking and copyright violations, are not responsible for what third parties do on their sites \cite{cda_1996}. This, combined with the broad and ambiguous policy of ``supporting free speech'' endorsed by many platforms has led to arrangements where much of the work of removing inappropriate posts falls on the shoulders of content creators and moderators \cite{Carlson2020}.

Our work contributes to the thread of content moderation research that focuses on techniques to counter online harm, including proactively preventing harm through norm-setting \cite{chandrasekharan2018norms,fiesler2018reddit}, increasing the cost of violation \cite{kiesler2012,grimmelmann2015virtues}, engaging in counter speech \cite{mathew2019thou}, and blocking certain keywords before they appear on the site \cite{seering:2017,jhaver2019automated}.
This line of research has previously explored how moderation teams for online forums rely on automated mechanisms to make moderation decisions \cite{jhaver2019automated,jiang2019Discord} and how computational approaches can be used to identify hate speech \cite{chandrasekharan2017bag}, pornography \cite{singh2016behavioral}, pro-eating disorder content \cite{Chancellor2017,feuston2020conformity}, and personal attacks, as well as high-quality comments \cite{park2016supporting}.
While machine learning approaches have frequently been proposed as a wholesale solution to proactively remove undesirable posts, they require gathering enough training data for rule violations, may not be adaptive to new kinds of infractions \cite{gillespie2018custodians,jiang2019Discord}, and their decisions may not be comprehensible \cite{jhaver2019automated}. 
Further, recent research suggests that algorithms designed by software engineers to detect hate speech or manage search functions show biases, e.g., they remove racial slurs but not gendered slurs \cite{noble2018algorithms,sap2019risk,Carlson2020}.
In this work, we examine creator preferences around simpler, yet more controllable, rule-based approaches to moderating comments, and we discuss the trade-offs with more sophisticated automated techniques.

Many platforms offer end-users moderation resources, such as the ability to report a post as inappropriate \cite{kou2021flag} or block a user from appearing on their profile \cite{cai2019effective,jiang2019Discord}.
However, these default moderation tools can fall short of addressing users' moderation needs.
For example, Matias describes how the prolonged neglect of moderation software on Reddit was so frustrating that it pushed moderators of 2,278 communities to join together in protest against the platform \cite{matias2016goingdark}.
To take another example,
a cross-platform study of queer female users showed that reporting to platforms failed to protect them from harassment and discrimination~\cite{duguay2020queer}.
Some critics have pointed out that the current systems' reliance on reports of dangerous or offensive content by users constitutes a reactive, too-little-too-late moderation approach that fails to protect the already-offended users \cite{ullmann2020quarantining,newton2019trauma,simon2019propaganda}.

Proactive approaches such as blocking accounts can be effective but can also be circumvented by motivated harassers creating new accounts~\cite{Mahar2018squadbox}. Some third-party tools \cite{party_2021,block_together_2021} address this weakness by allowing users to configure rules and share lists to auto-block accounts; we draw inspiration from these tools towards content-based moderation.
Creators on sites like YouTube, Twitch, and TikTok also have a larger suite of moderation tools than regular users. 
For example, Twitch creators can configure an automated moderation tool called Automod to select one of five levels (including the lowest level that turns off Automod) that affects how aggressively AutoMod prevents messages under four categories offered by Twitch: discrimination and slurs, sexual content, hostility, and profanity \cite{twitch_moderation_2021}.

\begin{figure*}
    \centering
    \begin{subfigure}[b]{0.475\textwidth}
        \centering
        \includegraphics[width=\textwidth]{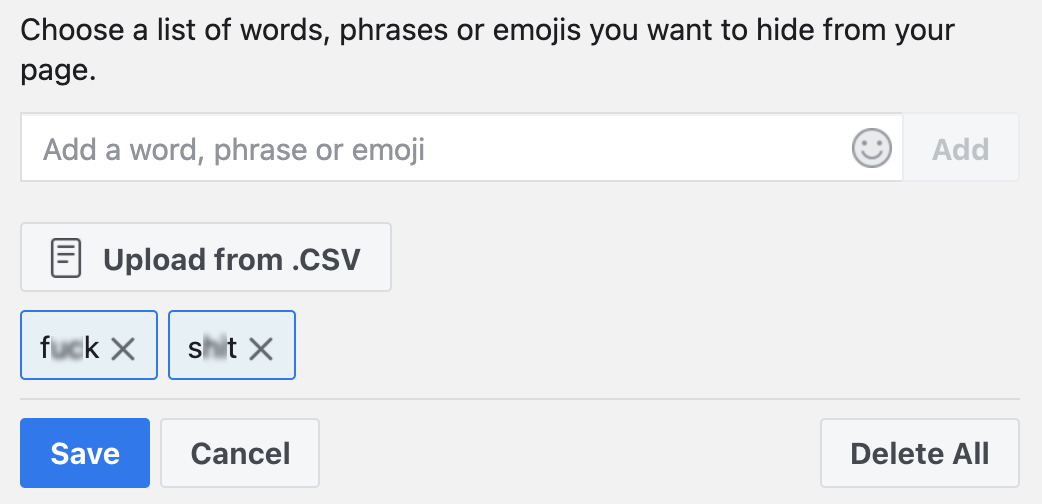}
        \caption {Facebook Page}    
        \label{fig:facebook}
    \end{subfigure}
    \hfill
    \begin{subfigure}[b]{0.475\textwidth}   
        \centering 
        \includegraphics[width=\textwidth]{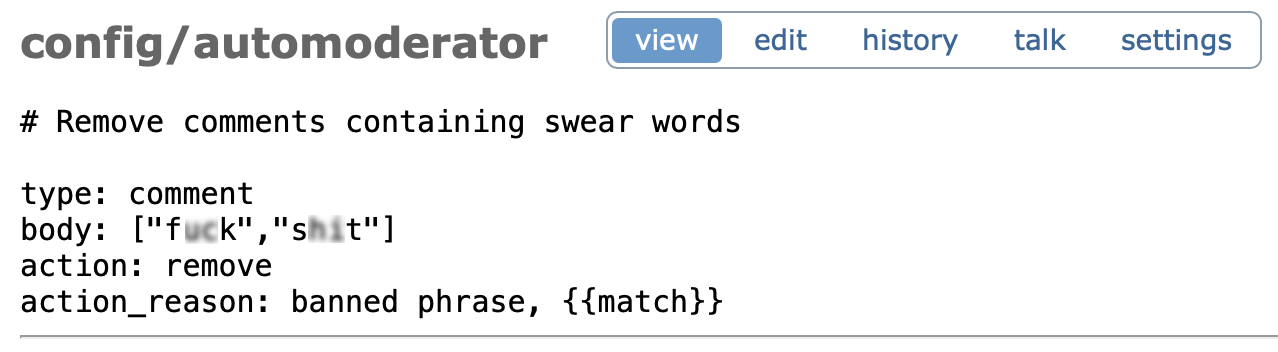}
        \caption {Reddit Automoderator}   
        \label{fig:reddit}
    \end{subfigure}    
    \vskip\baselineskip
    \begin{subfigure}[b]{0.775\textwidth}  
        \centering 
        \includegraphics[width=\textwidth]{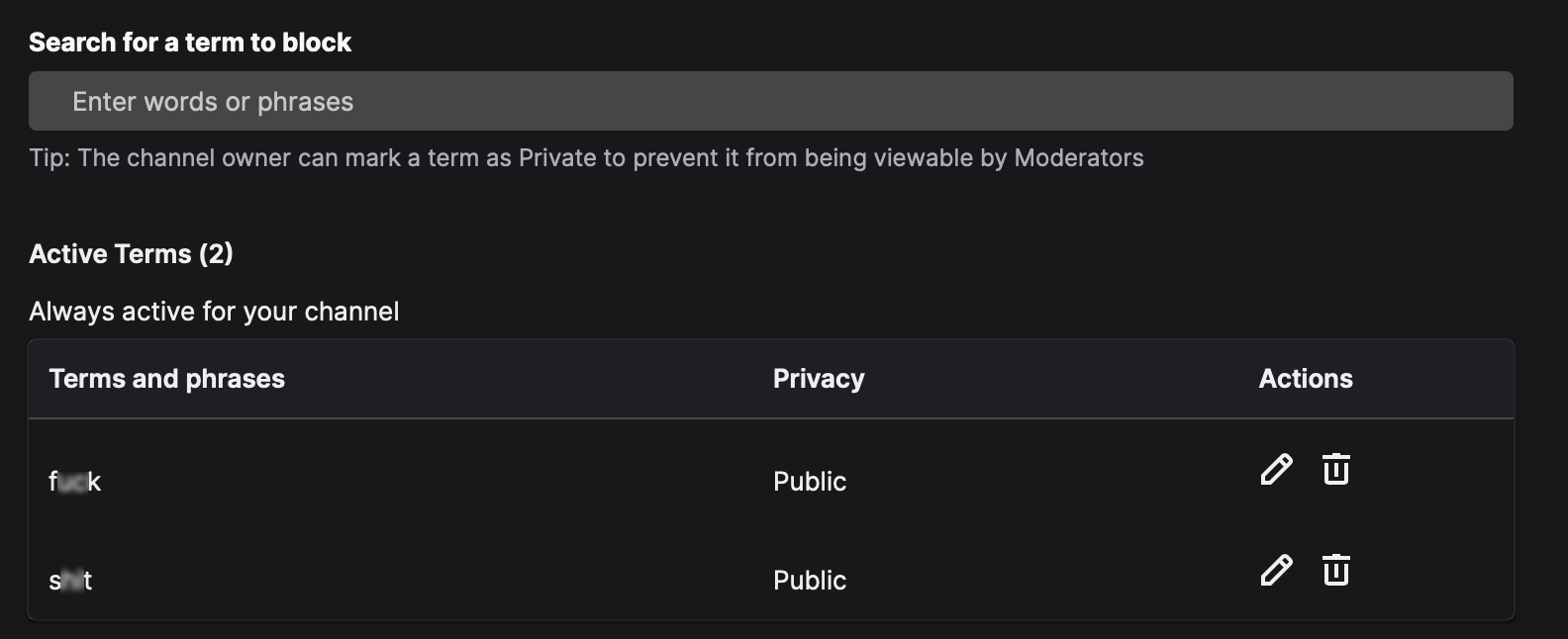}
        \caption {Twitch}    
        \label{fig:twitch}
    \end{subfigure}
    \hfill
    \begin{subfigure}[b]{0.675\textwidth}   
        \centering 
        \includegraphics[width=\textwidth]{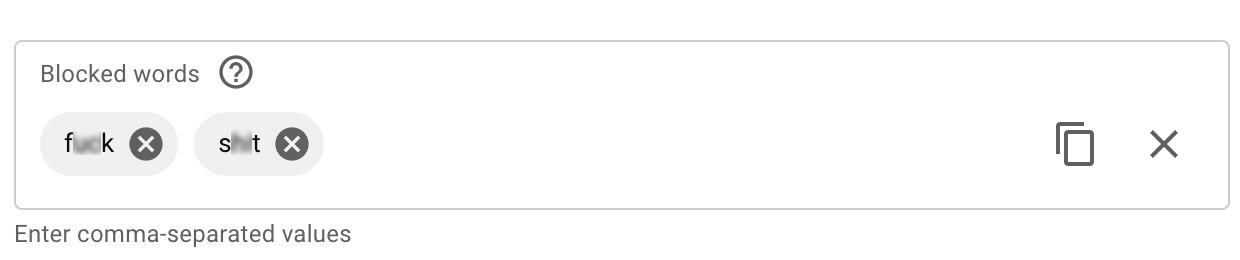}
        \caption {YouTube Studio}   
        \label{fig:youtube}
    \end{subfigure}
    \caption{Implementations of word filters on four popular platforms---Facebook, Reddit, Twitch, and YouTube. (a) This setting appears on Facebook Pages, which are public Facebook profiles that allow ``businesses, brands, celebrities, causes and organizations to reach their audience.'' \cite{fb_bc_2021}, (b) this community-specific setting is accessible only by the volunteer moderators of each Reddit community \cite{jhaver2019automated}, (c) this setting is available to Twitch creators on their \textit{Creator Dashboard}, a webpage that allows creators to organize content and improve safety preferences \cite{twitch_cd_2021}, and (d) this is available as part of \textit{Community Settings} on YouTube Studio, an official YouTube site where creators can grow their channel, interact with audiences, and manage earnings \cite{yt_studio_2021}.} 
    \label{fig:word-filters}
    \Description{Screenshots of word filter implementations on four popular platforms: Facebook, Reddit, Twitch, and YouTube.}
    
\end{figure*}

\subsubsection{Word filter moderation tools}
One of the most common types of end-user moderation tools, and the focus of this paper,  is \textit{word filters}, which lets users configure a list of blocked keywords. Once a keyword is configured in a word filter, any post containing that keyword is automatically removed or held for manual inspection.
Our review of moderation tools across popular social network sites found that word filters are offered by YouTube, Twitter, Facebook Pages, Reddit, Instagram, and Twitch, among others. 
Word filters are a subset of rule-based moderation tools, or tools that allow end-users to author generic rules over comments that can be more complicated than exact keyword matching. 
For example, Reddit moderators can configure automated removal of inappropriate comments using rules that specify \texttt{<regular expression, action>} pairs (see Figure \ref{fig:reddit}) in a subreddit-specific tool called Automod \cite{jhaver2019automated}.

As online communities or channels grow to a size that human moderation can no longer handle, rule-based moderation tools offer a hopeful solution for enacting moderation at scale. These tools can reduce the time-consuming work and emotional labor required of human moderators by automatically removing large volumes of inappropriate content.
However, we note that current word filter interfaces are relatively rudimentary---little more than text boxes for inputting comma-separated terms---with few features that incorporate interactivity, visualization, or affordances for information management (see Figure \ref{fig:word-filters}).
Sometimes they are configured using regular expressions, which are difficult to set up and require moderators to develop new technical skills.
Prior research has also highlighted the need for audit tools that provide moderators with greater visibility into the performance of such tools \cite{jhaver2019automated}. 
In this paper, we examine in detail the current use of word filter tools, identify the problems that creators face in using them, and contribute a system to address those problems.



\section{Needfinding Study: Motivations, Experiences, and Strategies} \label{sec:experiences}
We began with a needfinding study to understand the current moderation practices of content creators, including their strategies and challenges with using tools like word filters. 

\subsection{Method: Interviews with Content Creators}
\subsubsection{Interviews}
We conducted semi-structured interviews with 19 content creators on sites like YouTube, Twitch, and TikTok between October 2020 and April 2021. We first prepared a \textit{semi-structured interview script} to understand the moderation practices and word filter usage of interview participants. As we gained more understanding of the fundamental moderation issues, we continued revising our interview script to elicit deeper insights. We began by asking questions about general social media use for sharing videos to understand the context in which participants work. This included questions about the platforms used to upload content, how and why participants review comments on their content and engage with their audiences, and the types of inappropriate comments they receive. Next, we explored how and why participants began to use word filters, the changes they observed as a consequence, and the difficulties they faced in configuring these filters. Further, we queried about the design features and changes they would like to see incorporated into these word filters. 

Based on a few initial interviews, we began designing \textit{multiple alternative solutions} for serving three user needs: (1) capturing meanings of undesirable ideas, (2) understanding which posts are caught by each word filter configuration, and (3) specifying actions to take on word filters. For example, to capture meanings of undesirable ideas when entering a keyword, we presented two design sketches: (a) auto-suggesting similar phrases that participants may want to filter and (b) auto-suggesting entire categories of offensive keywords such as antisemitism and transphobia. 
To take another example, when setting up a word filter, we presented two ideas for configuration assistance: (a) showing examples of recent comments matching that filter and (b) showing descriptive statistics (e.g., number of comments, number of users, number of comments by the channel's moderators) of comments caught by that filter.

We described these solutions to participants and asked them to reflect on the solutions. Asking participants to contrast alternative solutions as shown above helped us elicit more creative and detailed descriptions. 
Most participants wanted a single tool offering multiple informational and visualizations features serving their different needs; we therefore attempted to design a tool that offers multiple features but does not cognitively overwhelm the user.
We constructed\textit{ preliminary design sketches} using the tool Figma to incorporate these features; we sought participant feedback on these sketches and continued revising the sketches based on  participant responses, particularly ensuring that the interface is easy-to-use. Our evolving design ideas and sketches led us to ask many additional questions addressing word filter categories that participants would like to have by default, and how participants felt about sharing their word filter configurations with other creators.

\begin{footnotesize}
\sffamily
\begin{table}[]
\caption{Demographic information of needfinding study participants. Here, `Platform Used' includes all the platforms where participants share the content they created. `Size' refers to the highest number of subscriptions participants received on their channel on any platform; for participants with size `N/A', the participant did not divulge the size of their channel. `Topic' refers to the main topical focus of creator's channels.}
\rowcolors{2}{gray!20}{white}
\resizebox{\textwidth}{!}{%
\begin{tabular}{llL{1cm}L{2.5cm}lL{2cm}lL{2cm}}
\toprule
\textbf{Sr. no.} & \textbf{Age} & \textbf{Gender} & \textbf{Occupation}                   & \textbf{Country} & \textbf{Platforms Used}              & \textbf{Size}      & \textbf{Topic}             \\
\midrule
P1               & 28           & Male            & Banker                                & USA              & YouTube, Reddit                      & \textgreater{}100k & American football          \\
P2               & 24           & Female          & Youtube Accounts Manager and Producer & Jordan           & YouTube, TikTok, Facebook, Instagram & 500-1k             & Family life             \\
P3               & 23           & Female          & Student                               & USA              & YouTube                              & 20k-50k            & School life                \\
P4               & 19           & Female          & Student                               & USA              & YouTube, TikTok                      & \textless{}500     & Music creation             \\
P5               & 20           & Male            & Student                               & India            & YouTube                              & \textless{}500     & Gaming                     \\
P6               & 21           & Male            & Student                               & Germany          & YouTube, TikTok                      & N/A                & Gaming                     \\
P7               & 24           & Male            & Student                               & India            & YouTube                              & \textless{}500     & Technology news and advice \\
P8               & 20           & Male            & Student                               & India            & YouTube, Reddit, Facebook, WhatsApp  & N/A                & Technology                 \\
P9               & 28           & Male            & Student                               & USA              & YouTube, Facebook, Instagram         & \textless{}500     & Satire                     \\
P10              & 29           & Nonbinary       & Freelance Creator                     & USA              & YouTube, Instagram                   & 20k-50k            & LGBTQ                      \\
P11              & 28           & Trans Female    & YouTuber                              & Australia        & YouTube                              & \textgreater{}100k & Gender identity            \\
P12              & 29           & Female          & Communty Manager                      & USA              & YouTube, Twitch                      & 5k-10k             & Miscellaneous              \\
P13              & 23           & Male            & Student                               & USA              & Twitch                               & N/A                & Gaming                     \\
P14              & 31           & Female          & Biologist                             & Brazil           & YouTube, Instagram                   & 10k-20k            & Science                    \\
P15              & 35           & Female          & Content creator                       & USA              & YouTube, Twitch                      & 500-1k             & Gaming                     \\
P16              & 27           & Female          & Content creator                       & UK               & YouTube, Twitch, Instagram           & 20k-50k            & Makeup                     \\
P17              & 35           & Male            & Podcaster                             & Brazil           & YouTube, TikTok                      & 10k-20k            & History                    \\
P18              & 29           & Female          & Customer Service Representative       & UK               & Twitch, Instagram                    & 500-1k                & Art                        \\
P19              & 28           & Male            & Student                               & Brazil           & YouTube, Instagram, Twitter          & 1k-5k              & Science                   
\end{tabular}
}
\label{table:participants}
\end{table}
\end{footnotesize}

\subsubsection{Participants} \label{sec:need_participants}
We adopted convenience sampling to recruit our subjects. We posted recruitment messages on social media sites and sent personal messages to participants we had interviewed in our previous studies. We sought to recruit a diverse set of participants, ensuring that our sample included creators with channels of different sizes who were focused on a variety of topics and reside in multiple countries.
Prior research has shown that BIPOC, LGBTQ+, and female users have experienced higher instances of online harassment and harms of content moderation \cite{haimson2021disproportionate,vaccaro2021contestability}. We therefore oversampled gender, racial, and sexual minorities since we expect that these groups would especially benefit from upgraded moderation tools.
Our selection criteria included choosing only those creators who had received at least a few dozens of comments on their channels so that they have more experience with the challenges of content moderation at scale.
We also selected only those participants who had either used YouTube Studio's \textit{Automated Filters} or other technical ways to manage comments on their videos since our interviews dealt with technical aspects of comment removals. 
Table \ref{table:participants} presents demographic information about our participants and the size and topical focus of their channels. To maintain privacy, we do not reveal participant names or channel details in this paper.

The interviews lasted between 75 to 90 minutes and were conducted through video chats using Zoom or Skype, depending on each participant's preference. Most participants used YouTube as their primary platform to upload content,  but some used other sites like Twitch and TikTok to share and stream videos. All interviewees received \$25 as compensation for their participation. 

\subsubsection{Analysis}
We fully transcribed our interview data and read them multiple times. One interview was conducted in Hindi; we translated it to English to simplify interview coding. 
Next, we applied interpretive qualitative analysis to all interview transcripts \cite{merriam2002}. This involved a rigorous categorization of our data to identify relevant patterns and group them into appropriate themes. We began with ``open coding'' \cite{charmaz2006}, for which we assigned short phrases as codes to our data. We conducted this first round of coding on a line-by-line basis to stay close to the data. Next, we engaged in multiple subsequent rounds of coding and memo-writing, conducting a continual comparison of codes and associated data. All authors discussed the codes and emerging concepts throughout the analysis and resolved conflicts through discussions. In later rounds, we began combining our initial codes into high-level ones, such as ``Prefers to handle comments automatically'' and ``Wants to see statistics.''
Once all authors agreed upon the codes, we distilled them into key themes and performed axial coding to deduce relationships. We next present these themes as our findings.

\subsubsection{Maintaining Participants' Privacy and Security}
Content moderation is a sensitive topic and creators, owing to their online public presence, are especially vulnerable to retaliation by internet trolls. Therefore, we took many important steps to ensure the privacy and security of our participants. We anonymized all our participants' identities in this manuscript and plan to continue doing so for all future presentations and demonstrations of this work. All comments and usernames shown on screenshots of our tool in this paper are synthetic. We also stored our interview recording and transcripts (for both need-findings analysis and the later user evaluation) on a secure, password-protected server.

\subsection{Findings: Moderation Concerns and Strategies}
We now describe content creators' experiences with receiving offensive comments, the moderation strategies they use to improve their content feeds, and their concerns with attaining balance in content moderation.

\subsubsection{Creators value engaging with user comments, but find frequent, undesirable comments disruptive} \label{sec:pub_undes}
%
All study participants indicated that they value reading user comments. Five participants whose channels focus on social justice issues or spreading public awareness about science and history pointed out that reading comments helped them realize how they are contributing to their audience. For example, P2 manages a YouTube channel that hosts videos about socially relevant issues and noted:
\begin{quote}
     ``\textit{I actually enjoy reading comments because they give you a sense of achievement, and it's really rewarding honestly, especially on videos where it's building awareness and seeing how people are reacting to that. For example, we did a video about gender equality, and it was really, really nice to see how people are reacting to that and their own experiences.}'' - P2
     
\end{quote}


%
%
However, viewing and engaging with comments becomes difficult when they are offensive, spam, or off topic. Echoing prior work, we found that participants felt particularly disturbed when these comments were harassing \cite{lewis2021we,uttarapong2021harassment}. For example, those from minority groups (users who were Asians, Blacks, LGBTQ, Muslims, female, or of Middle Eastern descent) reported experiencing identity-based attacks that severely distressed them.
Two participants also reported suffering from organized harassment or waves of negative comments that emotionally hurt them. P3 expressed concern about her audiences viewing offensive comments:

\begin{quote}
    \textit{``It feels almost kind of embarrassing for someone to say like, "You're ugly," or, "You're dumb," and then for other people to see that comment. I want to form a positive community, and if I don't control those sorts of language, then people will think it's okay, and then slowly the community will become a lot more negative overall.'' - P3}
\end{quote}

P16 noted that as channels become more popular and attract more viewers, the probability of receiving such offensive comments also increases.
In one case, such harassment caused the participant to have anxiety attacks, seek medical help, and subsequently stop posting videos on his channel. Female, transgender, Muslim, and BIPOC users were particularly bothered by receiving comments on their appearance. For instance, P14 talked about her experience of having a transgender co-creator on her science-based channel:
\begin{quote}
    \textit{``[We receive comments about] that friend of mine that is a trans woman about what is she doing, what is she wearing, why is she talking like that, why is she there? But she also is a scientist, she also has a PhD, and she is an amazing paleontologist.'' - P14}
\end{quote}

\subsubsection{Creators use a variety of strategies to moderate}
While nine participants stated that they often just ignored offensive or off-topic comments, others felt the need to engage in a variety of moderation practices to curb harassment both for themselves and their viewers. In the most extreme cases, participants disabled the comment section entirely, restricted the visibility of their account or video, or even removed the video. For example, P10 stated:
\begin{quote}
    \textit{``The only way to stop [the harassment] from happening was... to make my account private and not let other people who are trying to come at me follow me.'' - P10}
\end{quote}

More commonly however, participants elected to use their platform's built-in word filter tool to hold comments for review or block commenters. 
%
In addition, most participants also manually moderated their comments to some extent. Five creators replied to offensive comments to try and educate commenters and encourage better behavior. 
Some participants, especially those who streamed on the Twitch platform, used the help of volunteer moderators to manually moderate their comment sections as well. Those participants usually had a set of guidelines for their moderators to follow:
\begin{quote}
    \textit{''It's different per channel. What I think is acceptable or is unacceptable may be different to another creator so... you tell your moderators what is acceptable and what isn't.'' - P16}
\end{quote}

\subsubsection{Creators may receive backlash for their moderation practices} \label{sec:backlash}
While creators struggle most often with sufficiently moderating their comments, five participants reported receiving backlash in response to moderation practices perceived as too strict. They told us that  commenters reacted poorly if they felt that moderation was overly restrictive.  P3 noted:
\begin{quote}
    \textit{``I've seen posts on Reddit or even just on someone's Instagram page that's like, `Oh she's removing comments,' or, `This person is removing hate comments,' or, `Why's my comment not appearing?' It's kind of like accusatory.'' - P3}
\end{quote}

Moderation is seen as striking a balance between \textit{too much} and \textit{too little}. Under-moderation results in harassment and abuse in the creator's comments section. Over-moderation, though, can drive viewers away from the creator's channel, possibly costing the creator financially, or even cause commenters to harass the creator more, as Participant P11 pointed out: 
\begin{quote}
    \textit{``I think [the comments] would get caught in the word filter and then they wouldn't show up on the video. And so, people would think that I'm manually deleting them, which would then... encourage them to get more angry.'' - P11}
\end{quote}

 \textbf{Summary.} In summary, creators value engaging with user comments, but frequently encounter inappropriate comments that disrupt this process. To address this problem, creators try to educate offenders, restrict content visibility, and get assistance from volunteer moderators. They see content moderation as a balancing act between preventing abuse and retaining audiences.

\subsection{Findings: Current Use of Word Filters and Need for Improvements}\label{sec:experiences-filters}
Next, we focus on the content creators' use of word filters, one of the most prominent moderation tools available to creators on many platforms. We examine the current practices of using these filters and surface crucial needs that are not currently met by this tool. Specifically, we explore what additional informational elements and visualization features content creators would like to have in this tool, how they would benefit from more automation, and their attitudes towards sharing word filters they configure with other creators.

\subsubsection{Incorporating a greater degree of organization}
%
%
15 participants wanted to create different categories of word filters, such as racism, sexism, etc., to separately group different sets of inappropriate comments. For example, Participant P2 felt that having multiple categories would allow her to better organize and contrast comments caught by different categories as well as configure separate sharing preferences for each category. Participants P4 and P7 desired multiple categories because they wanted to configure different actions for each category. 
Five participants noted the utility of separating comments into distinct groups for reasons beyond moderation, such as easier review and search.
\begin{quote}
    \textit{``Well, we do it manually---we read every single comment that we get. And sometimes we get thousands and thousands of comments...
    So it would take so much less time and effort to just have them categorized into...a category where you can just see what sort of comments that you're getting and the percentages of good to not so great.'' - P2}
    
\end{quote}

\subsubsection{Assistance with configuring keyword spelling variants}\label{sec:need_variants}
When asked how they use these word filters, most of our participants noted that they configure them with curse words. 
However, eight participants experienced commenters evading and bypassing word filters to get their offensive comments read. This could involve using variations in the spelling of a slur or offensive word, spelling out offensive words phonetically or in a different language, or many other methods. 
%
%
To address these problems, nine participants wanted the ability to more easily configure similar words or creative misspellings. 
At present, this would require the  creator to either spend considerable time configuring all possible variations of a slur or to spend more time manually moderating and removing comments. When presented with a design where spelling variants are automatically generated when adding a slur to a word filter,  P10 responded:

\begin{quote}
    \textit{``It takes a lot of the burden of having to ideate all of this off of the creator...without this, that's a real emotional moment of labor of having to again, manually input all these things and create these really long block lists.'' - P10}
\end{quote}

\subsubsection{Understanding what different configurations would capture} \label{sec:need_conf_capture}
%
%
%
One of the most common problems that creators related about their use of word filters was that the filters catch many false positives. Frequently, built-in word filters flag innocuous, even valuable, comments and bury them in review folders that are full of genuinely offensive or spam comments. P2 said:

\begin{quote}
    \textit{``Sometimes, like in the last few months, I would post a video and then like an hour after, I would just go check the spam filter, and there were just be people talking about normal stuff...There's probably a bunch of nice comments caught in there but I don't really have the brain space to go back and look through them and approve all of them, so I just kind of leave them there, which sucks.''} - P11
\end{quote}

These issues resulted in three creators choosing to not use word filters at all. For example, due to the frequency of false positives, P15 pointed out:
\begin{quote}
    \textit{``I have [the word filter] disabled right now because it was being very excessive.'' - P15}
\end{quote}

In response to these issues and limitations, nine creators desired to see examples of matching comments when configuring keywords in word filters. They believed this would enhance both their understanding of how the word filter would perform and how to improve its performance. For example, P7 said:

\begin{quote}
    \textit{``[Seeing examples of matching comments] is interesting for me, because you can understand---because in a sentence, you can have multiple meanings of it, because even the word `alpha'... it's a smartphone name as well. So, if it is included in the sentence, you could mean a lot more because I can understand if they're talking about the phone or if they are slurring at me.'' - P7}
\end{quote}

\subsubsection{Importing built-in categories}
Black, female, and Muslim participants expressed a need for word filters to have a few default curated categories, like racism and sexism, that they can opt into so that they do not have to create such commonly needed categories from scratch. 
%
%
 P15 noted:
\begin{quote}
    \textit{``If you guys created some sort of like---`we've gathered together a panel of these five people to create our lists for this year and then every year it was edited'---right, because culture is constantly evolving, then I'd know that you guys have taken the time to select each of these people, and then I'm more likely to trust you guys as thought leaders in the space and use the lists.''} - P15
\end{quote}

\subsubsection{Getting an overview of word filters' actions}\label{sec:need_overview}
In addition to seeing what individual word filters would capture at the time of configuration (Sec. \ref{sec:need_conf_capture}), creators were also interested in reviewing the overall temporal performance of configured categories. For instance, twelve creators wanted to see a summary of actions taken by word filters, e.g., descriptive statistics about a filter's performance. P9 noted:


\begin{quote}
    \textit{``Obviously, something like adding word filters could potentially have a negative impact if 75\% of your viewers are saying phrase that you filtered out. And now because you have an automatic delete function, now your comments have dropped 80\% or something like that. So it probably would be cool to have statistics for stuff like that.''} - P9
\end{quote}

\subsubsection{Sharing word filters} \label{sec:needs_share}
Participants expressed diverse views on whether and how they would like to share their word filters with others. Four creators noted that they already use lists configured by other creators by manually copying and pasting them, but found doing so inefficient. 
Five participants also had a desire to collaboratively create word filters with other creators with similar interests.

\begin{quote}
    \textit{``A collaborative list would be nice, especially for female YouTubers because we have the most hateful comments, so there's some times that someone receives a bad comment with a word that another didn't have yet, but can appear sometimes, so collaboratively creating these lists would be great.''} - P14
\end{quote}

Two participants noted that they configure private or sensitive information such as their residential address in word filters to deter posters from doxxing them. Therefore, they would want to keep certain categories of word filters private. Three creators  worried that allowing everyone to share their word filters with one another would make them open to exploitation by bad actors or may proliferate poorly configured word filters that result in greater inaccuracy. P15 said:

\begin{quote}
    \textit{``If I know a particular creator,...and I already buy in and trust the way that they run their community and the way that they keep people safe within their community, then I am more likely to say, I know where this came from, I know there's likely reasoning behind most of it, and someone has been thoughtful about this...If it's just everybody always contributing, while that's probably easier to write new lists, it's probably not thoughtful enough to keep the gates closed to people who would abuse it.''} - P15
\end{quote}

\begin{footnotesize}
\sffamily
\begin{table}[]
\rowcolors{2}{gray!20}{white}
\caption{Design goals identified in our needs analysis. We also specify the Findings subsection from which each design goal arose.}
\begin{tabular}{clc}
\hline
\textbf{\#} & \textbf{Design Goal}                                    & \textbf{Connection to Findings} \\ 
\midrule
G1          & Incorporate a greater degree of organization            & 3.3.1                           \\ 
G2          & Include spelling variants of configured phrases            & 3.3.2                           \\ 
G3          & Preview expected effects while configuring word filters & 3.3.3                           \\ 
G4          & Offer importing of word filter categories               & 3.3.4                           \\ 
G5          & Show the effects of configured word filters             & 3.3.5                           \\ 
G6          & Allow sharing of word filters with other creators       & 3.3.6                           \\ 
\end{tabular}
\label{table:design-goals}
\end{table}
\end{footnotesize}

 \textbf{Summary.} We determined several design goals necessary for improving the usability of word filters. 
 First, participants noted that the current user interface for blocked words is a simple text box on most platforms, which does not meet their need for a greater degree of organization. They wanted the ability to configure multiple categories of word filters so that they could view and moderate different types of comments differently (G1). 
 Another issue with a comma-separated list of phrases is lack of organization around spelling variants of a phrase, which participants expressed that they often wanted to include (G2).
 Third, they wanted more visibility into the functioning of word filters, especially through observing examples of comments that would be caught by the filters they configure (G3). 
 Participants also wanted the ability to import existing categories of word filters to reduce the labor of initial configurations (G4). 
 They were eager to see statistics about word filters' actions over time (G5). 
 Finally, some participants wanted to share their word filters with other creators (G6).
 Table~\ref{table:design-goals} summarizes these design goals.
\section{FilterBuddy: A Word Filter Tool for Creator-Led Moderation}

As described in the previous section, our needs analysis enabled us to determine a set of design goals that would empower online content creators. With these goals in mind, we next designed and built \textit{FilterBuddy}, a system for creators to configure rule-based word filters. We developed FilterBuddy for YouTube, the platform of choice for many creators, after we observed that it contains only basic tools for configuring blocked words. However, FilterBuddy is a general framework applicable to any content sharing platform.
We next describe a scenario inspired by our subjects of how FilterBuddy can be used, followed by features and implementation of the system.

\begin{figure*}[]
\begin{tabular}{c}
\centering
  \includegraphics[scale=0.04]{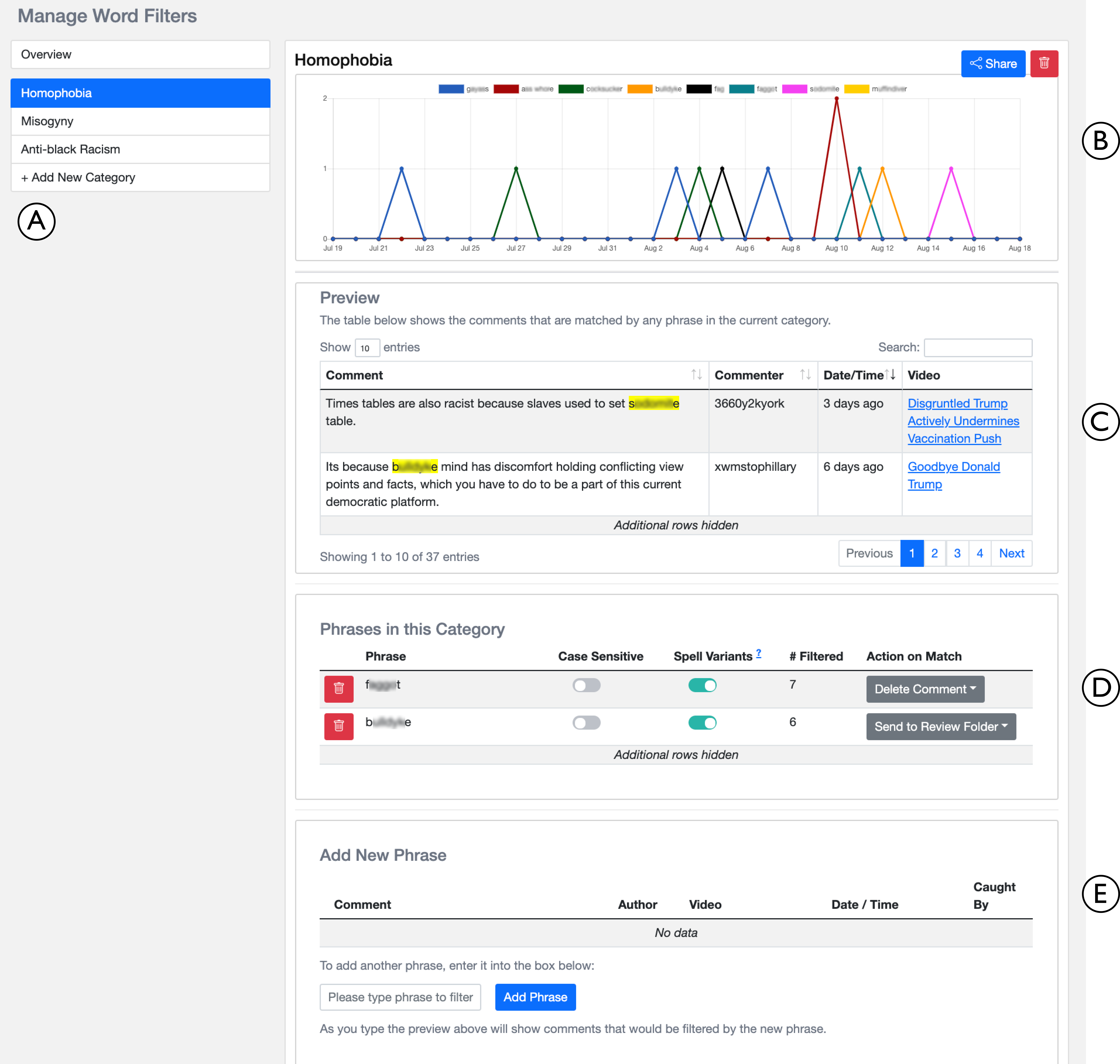}
\end{tabular}\\ [6pt]
\caption{A screenshot of FilterBuddy's category page showing (A) a sidebar with links to the overview page, each configured category page, and \textit{Add new category} page; (B) a chart showing the number of comments caught by each category phrase in the past month; (C) a paginated table previewing all comments caught by the category; (D) a table of phrases configured in the category with options to include/exclude spelling variants and determine action on match for each phrase; and (E) a section to add new phrases in the category. Note that we limit the number of table rows we show in all the figures for brevity.}
\label{fig:category_page}
\Description{A screenshot of FilterBuddy's category page showing (A) a sidebar with links to the overview page, each configured category page, and \textit{Add new category} page; (B) a chart showing the number of comments caught by each category phrase in the past month; (C) a paginated table previewing all comments caught by the category; (D) a table of phrases configured in the category with options to include/exclude spelling variants and determine action on match for each phrase; and (E) a section to add new phrases in the category.}
\end{figure*}

\subsection{User Scenario}
Isabella is a YouTube creator who likes sharing videos of herself cooking authentic Iranian recipes on her channel. 
She recently posted a video where she shared her spiced meat kebab recipe, which went viral and as a result, her channel suddenly receives thousands of new subscribers and comments.
While Isabella appreciates many of the encouraging comments and requests for sharing other Iranian recipes, she also receives a few comments that criticizes her looks and her clothing. Some other comments criticize her Iranian heritage while a few others are blatantly Islamophobic. Isabella likes to keep her channel conversations focused on celebrating Iranian food; therefore, she manually starts removing inappropriate comments. However, she quickly realizes that this manual process would be very time consuming. She also notices that many comments include the same or similar offensive words.

\begin{figure*}[]
\begin{tabular}{c}
\centering
  \includegraphics[scale=0.3]{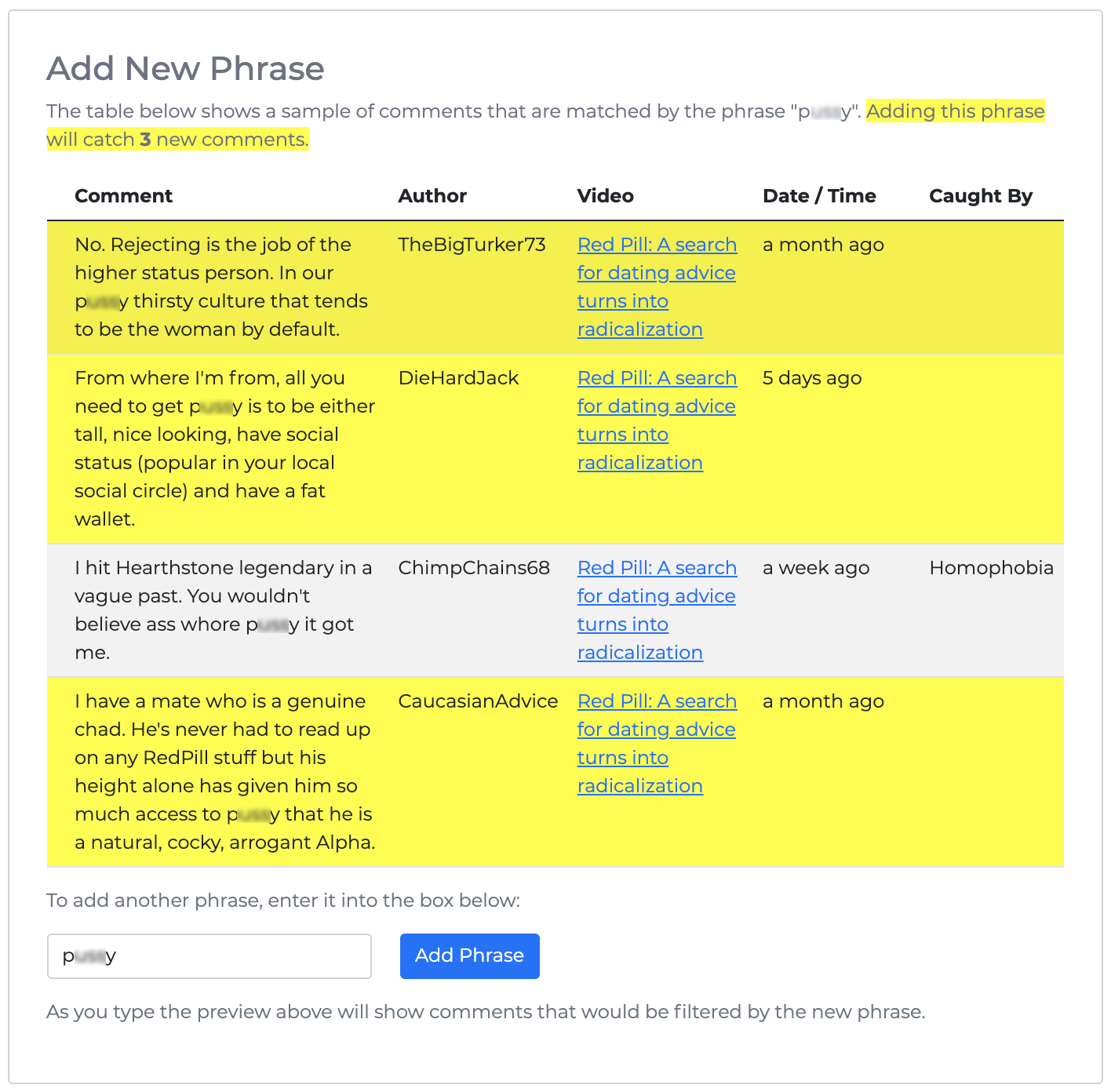}  
\end{tabular}\\ [6pt]
\caption{`Add New Phrase' section on the FilterBuddy Category Page. As the user types in a phrase, the comments caught by that phrase are auto-populated. Comments not already caught by any configured phrases have a yellow background so that they are easily distinguished.}
\label{fig:add_new_rule}
\Description{A screenshot of `Add New Phrase' section on the FilterBuddy Category Page. As the user types in a phrase, the comments caught by that phrase are auto-populated. Comments not already caught by any existing category have a yellow background so that they are easily distinguished.}
\end{figure*}

Looking to reduce her stress and time investment in removing undesirable comments, she logs in to FilterBuddy using her YouTube credentials. She notices that FilterBuddy already provides some importable categories (Figure \ref{fig:create-category}), and she chooses to import a category for \textit{Misogyny}. She is redirected to her new Misogyny category page (Figure \ref{fig:category_page}) that is pre-configured with 23 misogynistic phrases, each phrase listed alongside the number of comments on her channel containing that phrase (Figure \ref{fig:category_page}-D). She sets `Delete Comment' for some phrases and `Send to Review Folder' for other phrases as the action to take on matching comments. 
Noticing that her channel also contains anti-Islam comments, she creates a new category from scratch called \textit{Islamophobia}. When adding each new phrase to this category (Figures \ref{fig:category_page}-E, \ref{fig:add_new_rule}), FilterBuddy shows her all the prior comments caught by that phrase---reviewing these comments lets her make informed decisions about whether to keep or remove the phrase.

A week later, she logs in to FilterBuddy, and her home page (Figure \ref{fig:overview}) shows her a graph with the temporal trends of her categories of Misogyny and Islamophobia filters. 
She is glad to notice that the number of misogynistic comments reduced in the past week.
She also scans her most recent comments on this page and notices that some comments not caught by any existing category are politically divisive. Desiring to remove these, she creates a new category for \textit{Politics} to remove comments with political content.

\subsection{FilterBuddy Features}
We now discuss how the features of FilterBuddy serve the design goals identified in Section \ref{sec:experiences-filters}.

\subsubsection{Incorporating a greater degree of organization (G1)}
Instead of organizing around individual keywords like most word filter tools,
FilterBuddy organizes around \textit{categories}, where each category contains a list of phrases (see Figure \ref{fig:category_page}-A). This allows for greater organization when managing larger numbers of word filters. Users can also receive analytics about categories instead of every single phrase, which can help with understanding higher-level trends.

\subsubsection{Configuring spelling variants for each phrase (G2)} 
Each phrase in FilterBuddy is not just an exact-match filter but a regular expression under the hood. We provide two forms of per-phrase settings, including whether matching should be case-sensitive and if it should cover spelling variants (see Figure \ref{fig:category_page}-D). 
For example, `ABCD' is an acronym that is a South-Asian slur, which, if configured with cases-sensitive setting on, will be caught in comments only where the term occurs with the capitalized case. 
With the \textit{spell variants} setting turned on, a phrase will also match plurals and words with characters repeated. For example, configuring \textit{`shit'} as a phrase would also capture comments containing \textit{`shiiit'} and \textit{`shittt'}. In the future, we will investigate more comprehensive approaches to capturing spelling variants while retaining some customizability. By tying these settings to a single phrase, we also increase organization (\textit{G1}), as typically, users would need to write out every variant as distinct comma-separated keywords. 
FilterBuddy allows configuring the same phrase as a word filter in more than one category: when a word filter configuration is changed in one category, the same change also reflects for that filter in other categories.

\subsubsection{Previewing comments caught by a phrase (G3)}
FilterBuddy has a dedicated section `\textit{Add New Phrase}' (Figure \ref{fig:category_page}-E) in each category page for users to add new phrases. As the user types in a new phrase, FilterBuddy will interactively show a preview of all previously published comments that would be matched by that phrase (see Figure \ref{fig:add_new_rule}). This preview also shows whether these comments have not already been caught by another phrase 
and reports the total number of uncaught comments at the top. 
We note that this feature may inadvertently expose users to abusive comments they receive. However, by letting users understand the contextual details of how particular keywords are used on their channel, it can help them make more informed decisions on how to automate moderation and see fewer of these comments in the future. We expect that creators will need to review only a few comments to make such decisions and once set up, more accurate configurations will better serve creators' needs (see Sec. \ref{sec:backlash}) over a longer duration.

\begin{figure*}[t]
\begin{tabular}{c}
\centering
  \includegraphics[scale=0.3]{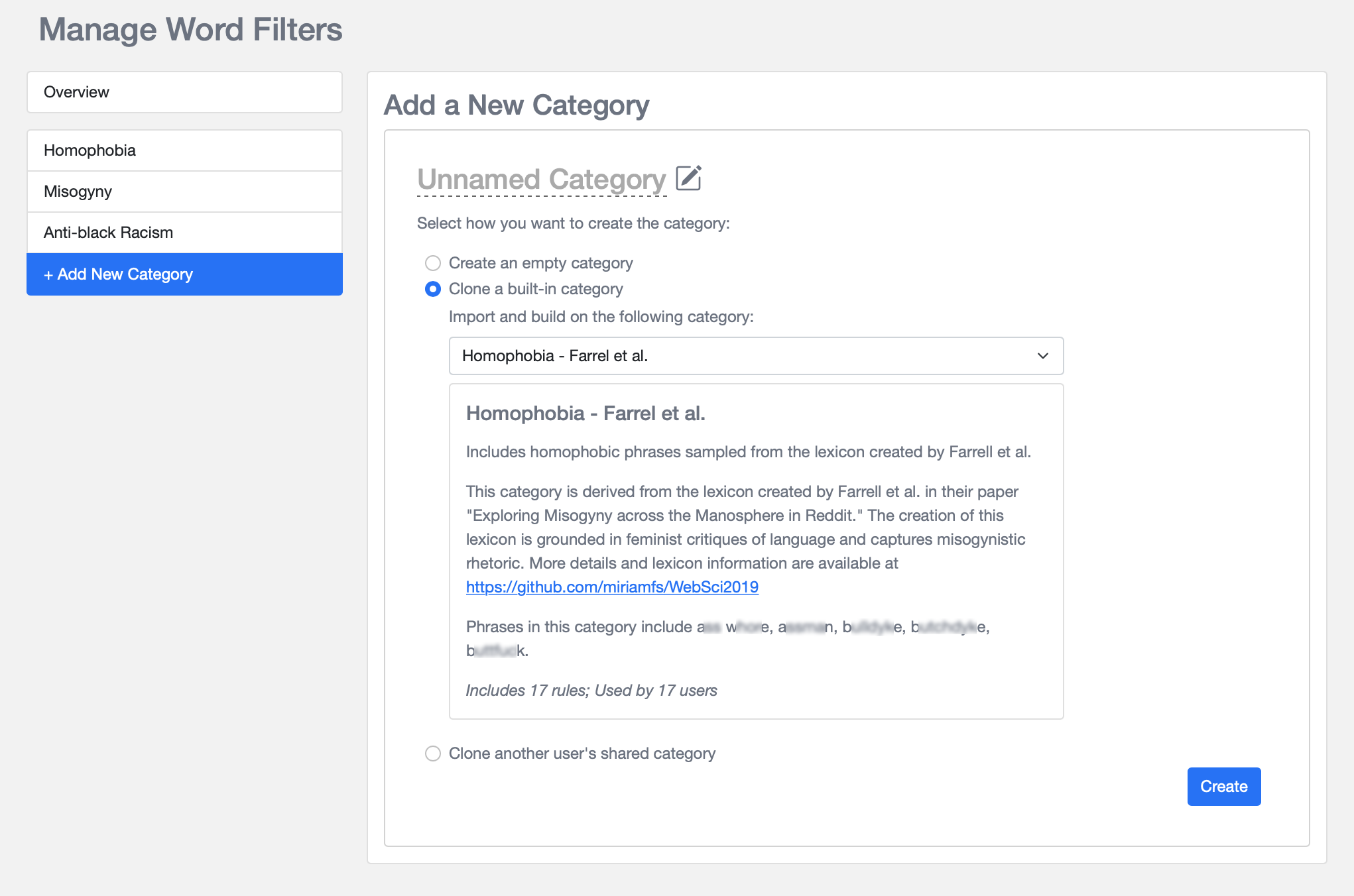}
\end{tabular}\\ [6pt]
\caption{FilterBuddy's `Add a New Category' Page. Users can either create an empty category, import one of the built-in categories, or clone a category shared by another user to quickly set up their configurations. We show here the details of the built-in `Homophobia' category selected in the dropdown.}
\label{fig:create-category}
\Description{A screenshot of FilterBuddy's `Add a New Category' Page. Users can either create an empty category, import one of the built-in categories, or clone a category shared by another user to quickly set up their configurations. We show here the details of the built-in `Homophobia' category selected in the dropdown.}
\end{figure*}

\subsubsection{Offering importable categories (G4, G6)}
Users can click `\textit{+ Add New Category}' in the FilterBuddy sidebar (Figure \ref{fig:category_page}-A) to create a new category. This redirects them to a page (Figure \ref{fig:create-category}) where they can name and create an empty category.
FilterBuddy also offers a set of built-in categories to make it easier for users to get started with creating word filters. 
Currently, FilterBuddy has the following importable categories: \textit{Homophobia}; \textit{Physical Violence}; \textit{Sexual Violence}; \textit{Pejorative Terms for Women}; and \textit{Anti-Black Racism}. We manually collected these lists of terms from searching through academic and other public resources.
This selection of categories was especially aimed at addressing the identity-based attacks that our female, LBGTQ, and black participants reported in our needs analysis.
For each of these built-in categories, we show an information box containing the following information:
\begin{enumerate}
    \item A brief description of the category
    \item The authors (website, organization, or researchers) that developed the category
    \item The number of rules contained in the category
    \item Examples of rules in the category
    \item The number of users who have imported this category
\end{enumerate}

After importing a category, users can then customize their copy of the category just as if they made it from scratch.

Finally, on this page, users can also import a category that has been shared with them by another user on the FilterBuddy site (\textit{G6}). For the purpose of our user study, we pre-loaded the dropdown list with categories that we had authored as a demonstration of this possibility; in a deployment in the wild, the interface would be populated by categories made by friends of the user. 
We leave to future research a deeper exploration of how to design sharing features such that users can choose who to share each category with (either a selected group of friends or everyone) and how they can use that category (importing versus collaborating). We see file sharing permissions on sites like Google Drive and collaboration models like Git as promising early models for designing FilterBuddy's sharing features.

\begin{figure*}[t]
\begin{tabular}{c}
\centering
  \includegraphics[scale=0.4]{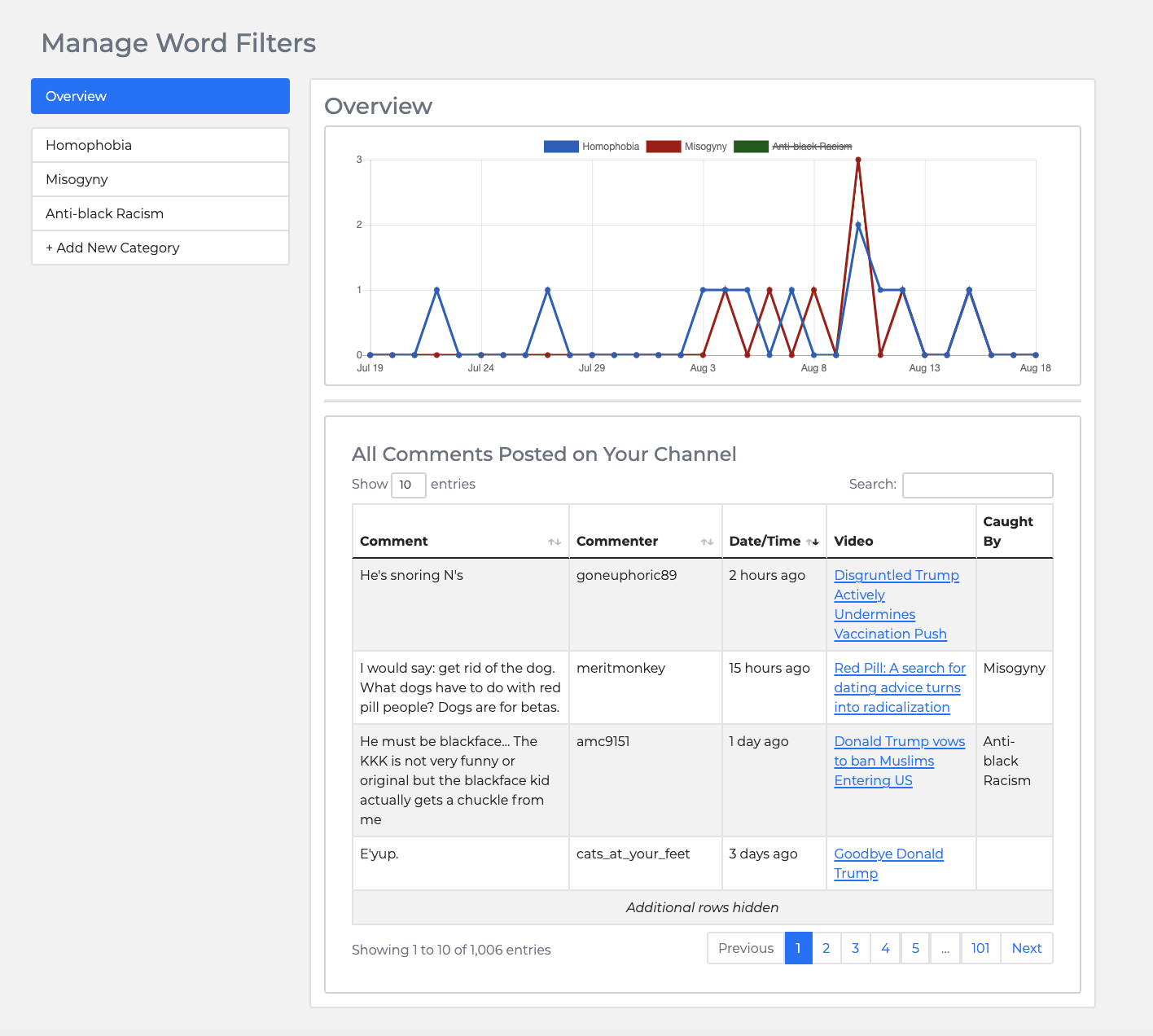}
\end{tabular}\\ [6pt]
\caption{FilterBuddy's Home Page. It shows (a) a time-series graph for number of caught comments aggregated at the category level and (b) a paginated table of all comments posted on the user's channel. }
\label{fig:overview}
\Description{A screenshot of FilterBuddy's Home Page. It shows (a) a time-series graph for number of caught comments aggregated at the category level and (b) a paginated table of all comments posted on the user's channel. }
\end{figure*}

\subsubsection{Showing a time-series graph and table of comments (G5)}
On its home page (Figure \ref{fig:overview}), FilterBuddy offers an overview of how different categories are functioning recently. This includes a graph that shows the number of comments caught by each category over the past thirty days. It also shows a table with all comments posted on the user's channel ordered by recency, along with sorting and searching capabilities. 
The table shows for each comment the category, if any, that caught the comment. Combined, these features help users quickly understand how their word filters are operating. On the category page, FilterBuddy presents a similar graph (Figure \ref{fig:category_page}-B) with the phrases that caught the most comments from the category.
There is also a table (Figure \ref{fig:category_page}-C) for each category that shows all comments caught by that category, with the relevant phrases highlighted.

\subsection{System Implementation} 
FilterBuddy is a Django web application. It uses a MySQL database to store user data.  Users log into FilterBuddy using their YouTube credentials. After they log in for the first time, FilterBuddy uses the YouTube API to retrieve all public comments posted on their channel and saves it to the database. 
As new comments get posted on YouTube, FilterBuddy periodically checks for and retrieves new comments.
Next, depending on how users configure their word filters, FilterBuddy takes actions on new comments as they are retrieved, such as removing caught comments, blocking users, etc., again using the YouTube API. These actions are reflected not just in FilterBuddy's local database but also on the creator's YouTube channel; FilterBuddy is not just a simulation but a tool designed to be actively used by creators. FilterBuddy does not currently support bulk or retroactive actions on older comments though this may be incorporated in the future.
We have released FilterBuddy's code on Github as open source, and we plan to publicly launch FilterBuddy as a free hosted service in 2022. 
\section{Evaluation} \label{sec:evaluation}
We introduced different features of the FilterBuddy tool and its workflow to eight YouTube creators (referred to as U1-U8), and then asked participants to explore the tool and author filters over their comments while providing qualitative feedback.  In this section, we discuss our methods and findings from this user study.


\begin{footnotesize}
\sffamily
\begin{table}[]
\rowcolors{2}{gray!20}{white}
\caption{Demographic information of user study participants.}
\begin{tabular}{cccccccc}
\toprule
\textbf{Sr. no.} & \textbf{Age} & \textbf{Gender} & \textbf{Country} & \textbf{Platforms Used}  & \textbf{Previously} & \textbf{Size}      & \textbf{Topic} \\
\midrule
U1               & 23           & Male            & USA              & YouTube, Twitch          & P13                 & N/A                  & Gaming         \\
U2               & 20           & Female          & USA              & YouTube, TikTok          & P4                  & \textless{}500     & Music creation \\
U3               & 24           & Female          & Turkey           & YouTube, Twitch          & -                   & \textless{}500     & Gaming         \\
U4               & 24           & Female          & USA              & YouTube, Twitter, Twitch & -                   & \textgreater{}100k & Makeup         \\
U5               & 29           & Female          & Germany          & YouTube                  & -                   & 10k-20k            & Academia       \\
U6               & 24           & Female          & USA              & YouTube                  & P3                  & 20k-50k            & School life    \\
U7               & 31           & Female          & Germany          & YouTube, Instagram       & P14                 & 10k-20k            & Science        \\
U8               & 29           & Female          & USA              & YouTube, Twitch          & P12                 & 5k-10k             & Miscellaneous 
\end{tabular}
\label{table:user-study}
\end{table}
\end{footnotesize}

\subsection{Methods}
We reached out to the same participants we interviewed during our needfinding study
and also conducted an additional round of social media recruitment. 
Five participants were previously involved in our earlier study, and we also gained three new participants.
Similar to our recruitment criteria in the prior needs analysis (Sec. \ref{sec:need_participants}), we oversampled gender, racial, and sexual minorities. We ensured that our participants had received at least dozens of comments on their videos, and they were familiar with settings available to automate content moderation. 
Table \ref{table:user-study} details the demographic information about our participants and the size and topical focus of their channels.
Each session was conducted online via Zoom for 45-90 minutes, recorded, and transcribed.
Before each session, we sent the participant details of our project and offered to answer any questions they had. 
All participants again received \$25 as compensation for their participation. 

We began each user study session by providing a brief overview and goals of FilterBuddy.
Next, we asked the participant to log into FilterBuddy using their YouTube credentials. During this step, FilterBuddy automatically retrieved all public comments posted on the participant's channel and populated the FilterBuddy interface. We then asked participants to explore the interface and create new word filter categories, either starting from scratch or importing a built-in or shared FilterBuddy category. 

After authoring word filters, the FilterBuddy  graphs and tables become populated with the comments posted on the participant's channel as if those filters had been turned on for the last thirty days. 
This was necessary to demonstrate the tool within the time period of a user study session; otherwise, participants would need to turn on the tool for an extended period of time.
Importantly, we chose in this study to not actually carry out any actions arising from any of the word filters that participants created, so as to minimize disruption to our participants and encourage freer exploration.
For the study, the tool only retrieved comments from participants' channels that were publicly readable, and we deleted any user data collected by the tool after their user study session was over.

We asked participants to ``think-aloud'' \cite{lewis1982using,van1994think} as they tried different FilterBuddy features, expressing their likes and dislikes and describing additional features they would like to have. We asked them to reflect on how they would configure FilterBuddy if they were actually using it to moderate their channels and imagine themselves coming back to this tool after having it on for thirty days.
%

We again conducted interpretive qualitative analysis \cite{merriam2002} on the transcripts of the user study sessions.
We also referred to the code book we had used for our needs analysis since it helped us associate our design goals with participants' evaluation of features designed to address those goals.
We now present the prominent themes we observed during this process.

\subsection{Findings}
\subsubsection{Participants emphasized control over greater automation} \label{sec:evaluation-automation}
All participants appreciated that FilterBuddy takes care of some spelling variants of phrases and had ideas for additional variants they wanted FilterBuddy to capture.
However, beyond common heuristics such as substituting `e' with `3', most participants were wary about more sophisticated tools for automatically learning what content to filter.
Instead, all study participants expressed a desire to retain control over and visibility into the operations of any moderation tool they adopt. 
U3 expressed concern about the accuracy of machine learning tools and wanted to see their execution and results before adopting them. In a similar vein, U6 said:

\begin{quote}
    \textit{``I would prefer to just do it myself manually. I almost wouldn't want that power taken away from me because moderation is so personal and difficult and very sensitive, so I think I would want complete control over it.'' - U6}
\end{quote}

However, participants had ideas of ways that automation could be introduced but without a loss of control. For instance, four participants wanted automated suggestions of new phrases to add to their existing categories, with the ability to choose whether to incorporate them. 

\subsubsection{Importable categories were seen as a powerful way to reduce toxic content}
Participants noted that it can be challenging to spontaneously think of phrases to configure, and they often learn about candidate phrases to block only after they see them appear in their comment feed. U6 also found it discomforting to explicitly type out offensive phrases like the N word. Therefore, most participants listed the ability to import built-in categories as their favorite feature of FilterBuddy. For example, U4 said:

\begin{quote}
    \textit{``Having the base list is just perfect for the creator, because they don't even have to like spend the time, just like brewing over what words might be used towards them, which is like the most messed up thing to have to think about in the first place, so having a tool like this is just amazing. Honestly, it's a lifesaver!'' - U4}
\end{quote}

Five participants noted the potential of using built-in FilterBuddy categories to broadly improve the content quality on platforms like YouTube and desired to see their widespread adoption.
 U8 suggested a new FilterBuddy feature that lets users propose additional phrases to include in built-in categories, so that they continue to respond to the evolving needs of marginalized users.
U8 also recommended that FilterBuddy encourage creators to import categories such as racism and homophobia to build more inclusive channels:

\begin{quote}
    \textit{``The creator is a cultural lead and, if you allow stuff that, oh it doesn't apply to me but like it's still gross, you're kind of defining that you allow that gross behavior, so some little reminder of like `even if you're not black, consider using this category' would be great.'' - U8}
\end{quote}



\subsubsection{Participants wanted defense mechanisms to prevent false positives}\label{sec:eval_defense}
Our participants showed a keen awareness of how overly-broad FilterBuddy configurations can create false positives and drive away users.
 U6 expressed concern that spelling variants of  phrases might  unintentionally include some acceptable phrases and suggested that users be shown all English dictionary terms that the spelling variants include and be able to make exceptions for those terms. 
 U3, who runs a Turkish YouTube channel, worried that configuring some English terms might unintentionally catch some acceptable Turkish terms; therefore, she appreciated that the system lets users toggle on/off the spelling variants of phrases. U5 wanted to build more complex rules that filter comments containing multiple specified phrases.

Participants appreciated the analytics and visualization features of FilterBuddy that helped them reduce false positives.
They liked the ability to scan through and edit the specifics of all phrases and their corresponding actions in each category.
All participants noted that seeing a preview of caught comments when adding a new phrase helped them make more informed decisions about setting up their filters: 

\begin{quote}
    \textit{`` I like that you can kind of check for how bad something might be here before you add it, I think that's pretty cool!'' - U8}
\end{quote}

\begin{quote}
    \textit{``I think that this is actually very helpful because it helps you understand how many of your word choices are incredibly common. For instance, if you were to use a word that's a super common phrase and this table shows that a lot of people are going to end up using it generally in any comment, that helps me understand like how much I should narrow it down compared to what I was doing.'' - U4}
\end{quote}

Participants also valued seeing other metadata about caught comments. For instance, U6 considered adding a swear word as a phrase, but, when seeing that the preview of caught comments for that word included amusing comments posted by her friend, she decided against doing so. Four participants noted and appreciated the visual distinction of seeing which comments were not already caught by existing phrases.

Importing built-in FilterBuddy categories or those developed and shared by other creators was seen as a way for false positives to creep in.
Participants noted that they would like to personalize and maintain their own copy of the imported categories. If new phrases were subsequently added to the original categories, they wanted them  to appear as suggestions to add on their category page rather than having them automatically added:

\begin{quote}
    \textit{``My ideal would be like, `Hey, changes were detected to this one that you copied from. Do you want to keep those, like, do you want to merge the changes in to your list - yes or no'.'' - U8}
\end{quote}

This focus on reducing false positives is in line with creators' desire to maintain a balanced approach to content moderation we discussed in section \ref{sec:backlash}. Indeed, we observed that creators who wanted to grow their channel's subscriptions were more wary of having user comments incorrectly removed. This is in contrast to Reddit moderators' focus on minimizing false negatives as opposed to false positives when using Reddit Automod, as observed by Jhaver et al. \citep{jhaver2019automated}. One explanation for this contrast could be that Jhaver et al. \citep{jhaver2019automated} studied large Reddit communities whose moderators were more concerned with reducing inappropriate comments than growing their subreddits.

\subsubsection{Participants wanted to use FilterBuddy for more than just removing offensive speech}\label{sec:eval_custom}
While every participant wanted to set up categories to address identity attacks, such as sexism, racial slurs, homophobia, and Islamophobia, other use cases were also frequently brought up.
U4 and U8 said that they would create a category containing phrases related to their personal information, such as social security number and address; they intended to use this category to preemptively prevent doxxing. 
U6, a female Asian creator, expressed a desire to set up a category to filter out comments that remark on her appearance using phrases such as `ugly', `fat', and `unattractive.' 
Five participants also discussed creating categories to remove off-topic, especially politically charged comments.

All participants noted the utility of FilterBuddy as a tool to filter comments in different buckets and better understand their audiences. Four participants wanted to set up a category to capture comments with positive emotions but not perform any action on them. For instance,  U7 wanted to set up categories for positive and negative phrases and visualize how the volumes of comments caught by both categories change over time. Participants also shared a number of creative categories and use cases that we had not anticipated. For example, U4, who operates a gaming channel, said:

\begin{quote}
    \textit{``I would probably make a self-promotion category because I know a lot of times you'll have people being like, `Oh, subscribe to my channel instead' or whatever, because it's not quite a hate comment, it's just something that's kind of frowned upon. Critical or constructive criticism comments would probably have like its own separate category. Maybe like a collab category so if anyone wanted to say, `Oh, you should play with me sometimes,' something like that, then it just filters those. Maybe game recommendation because I definitely play a lot of games, so say a phrase like `You should play...', I'd put that in there; it gives you an idea of how often are people saying I should play something else.'' - U4}
\end{quote}

\subsubsection{Participants were keen to engage in community building with other creators by leveraging the sharing feature} \label{sec:eval_share}
Participants reacted positively to the idea of letting other creators benefit from their word filters, especially when context-specific moderation is required, and multiple creators have similar moderation needs. For example, U3 noted that FilterBuddy's built-in categories might not cover all moderation requirements; therefore, letting users share categories may fill essential gaps:

\begin{quote}
    \textit{``Sharing is really useful because you will not be able to think of everything that is a problem in the Internet. So in Turkey, there may be things that are not a problem in America or in Korea or Europe. So, if a Turkish YouTuber designed her categories to be exactly what other Turkish YouTubers would want, it will be really awesome to be able to share her categories and will be really time saving, too.'' - U3}
\end{quote}

 U8 wanted the ability to share not just the phrases she set up in any category, but also the actions configured for each phrase, and information about whether she turned on the case-sensitive and spelling variant setting for that phrase:

\begin{quote}
    \textit{``I think, as long as they know they can change things themselves, it is not a bad idea to have it shared the way that creator set it up because if I were passing it on to somebody else, I want to make it as seamless as possible....'' - U8}
\end{quote}


\subsubsection{FilterBuddy was seen as a powerful tool to help content creators}
Overall, participants were excited to use FilterBuddy and appreciated its utility in not just removing unwanted content but also in helping them understand topical trends in their comments.
Six participants appreciated the time-series charts for different categories and three of them expressed an interest in using the graph spikes as clues to further filter and investigate comments causing those spikes. 
U6 predicted that FilterBuddy would reduce the emotional labor of removing inappropriate comments. 
U8 anticipated that FilterBuddy would deliver substantial time savings for creators:

\begin{quote}
    \textit{``I think it is empowering and also good for people's growth because, whether you're a big YouTuber with you know, three editors and a manager or if you're a small creator who just makes videos in their free time, that's still time you don't have to spend on manual moderation.'' - U8}
\end{quote}

Participants also commented on benefits for marginalized groups since such groups receive disproportionate number of undesirable comments. For instance, U8 felt that black and female content creators would especially benefit from using this tool because of its proactive support to address sexism and racism.
U2 noted that creators who have been targeted by commenters in the past would be especially motivated to use this tool to preemptively and automatically address online attacks.
Overall, participants did not find it cognitively challenging to understand how FilterBuddy works and each participant appreciated the confluence of multiple useful features that it offers. 

\begin{quote}
    \textit{``If you want something generally family friendly, this is just like a tool to help that happen. It's a one stop shop to solve moderation issues and manage your comments.'' - U1}
\end{quote}

\section{Discussion}

\subsection{Design Implications}

Drawing on both our system evaluation (Sec. \ref{sec:evaluation}) and needfinding results (Sec. \ref{sec:experiences}), we present the following design implications that navigate different trade-offs we identified between creator goals.

\subsubsection{User control when using automation}
One of our key findings from both our needfinding (Sec. \ref{sec:need_conf_capture}, \ref{sec:need_overview}) and system evaluation (Sec. \ref{sec:evaluation-automation}) is the extent to which content creators value retaining control and having visibility over their moderation operations. As noted earlier, study participants preferred to retain granular control over the actions associated with each configured keyword and to specify whether they want to include spelling variants (Sec. \ref{sec:need_variants}). 
However, additional control can incur a trade-off with manual effort.
Turning on the spelling variants setting has the advantage of reducing the number of separate rules that creators would need to set up for terms with similar meanings. On the other hand, turning this on also reduces visibility into exactly which keywords are currently set up to trigger the configured actions and increases the chance of obtaining false positives. 
Participants were also wary of catching false positives with greater automation, which made them hesitant to use machine learning tools to remove inappropriate content (Sec. \ref{sec:evaluation-automation}). This highlights the importance of reducing burden on users with sensible defaults in place that still let them retain granular oversight.

Prior research has surfaced similar tradeoffs in other domains involving human-computer integration \cite{farooq2016human} and explored questions such as the type of tasks that should be delegated to AI, the humans' preferences for what role AI should play, and ways to evaluate the division of labor between humans and automated systems \cite{amershi2019guidelines,feuston2021putting,mackeprang2019discovering,lubars2019ask}.
Evaluating a dataset of 100 human-machine collaboration tasks, Lubars and Tan found a disinclination towards ``AI-only'' designs and a preference for machine-in-the-loop designs \cite{lubars2019ask}, a finding echoed in our study. They also found trust in AI system to be an important factor in shaping human preferences of optimal human-machine delegation---this is in line with our participants' preferences for easily controllable, transparent regular-expression based moderation tools.
Mackeprang et al. highlight the challenges of finding the right-level of algorithmic support that achieves the compromise between minimizing human effort and maximizing system performance \cite{mackeprang2019discovering}. Their methods of systematically defining and evaluating different levels of automation could serve as a template for further development of AI tools for content creators.

We see more advanced AI approaches playing a critical role in further strengthening the value of tools like FilterBuddy. 
For example, training data based on the labels generated by configured word filters can be used to train ML models to obtain phrase suggestions for users to add to their existing categories. 
Further, collaborative filtering techniques \cite{schafer2007collaborative} can be employed to suggest creators additional categories to configure.
However, for any ML additions to FilterBuddy, developers should remain mindful of the creators' need to maintain agency over their tools.
More broadly, we carry forward the arguments made elsewhere that when integrating ML in the workflow of human-managed systems, designers and developers should carefully attend to the needs of their users that go beyond traditional metrics, such as precision and recall of ML models \cite{davis2006relationship}. 

\subsubsection{Designing features to better understand and be responsive to audiences}
The study participants' enthusiasm about developing custom categories (Sec. \ref{sec:eval_custom}) and visualizing resulting graphs (Sec. \ref{sec:eval_defense}) is indicative of content creators' need to better understand their audiences. 
This suggests that creators would benefit from advanced analytic and visualization tools that facilitate discovery of unusual temporal and topical trends in their comments.
For example, creators may value receiving  notifications of unusual spikes in posting activities or the use of toxic language on their channels.
We were also surprised to find that participants wanted to use our tool for more than moderating out unwanted content (Sec. \ref{sec:eval_custom}).
It indicates that creators may value new, creative ways of understanding their audiences where they have the power to specify which phrases or activity they want to monitor.

As content creators invest substantial amounts of time and emotional labor to gather and sustain an online audience \cite{chan2019becoming,gray2019ghost}, we also found that fears over audience backlash and exodus in response to excessive moderation are a key factor in creators' moderation strategies (Sec. \ref{sec:backlash}).
While the configuration assistance and performance visualizations provided by FilterBuddy are helpful in setting up rule-based moderation (Sec. \ref{sec:eval_defense}) that reasonably balance these goals, creators still need to exert significant care in their decisions about which filters to configure. Tools could additionally incorporate features such as ways for creators to have a trial run of a new filter or be able to revert a filter that was instated and bulk restore comments that were removed.



\subsubsection{Sharing and collaborating on word filters}
Study participants expressed an interest in sharing word filters with one another to let other creators benefit from their configurations (Sec. \ref{sec:needs_share}, \ref{sec:eval_share}). However, they were also concerned that sharing these word filters might raise the possibility that these configurations are leaked, rendering them vulnerable to exploitation by bad-faith actors who use creative misspellings to bypass word filters (Sec. \ref{sec:needs_share}).
Tools will need to consider how users can signify trust to each other when deciding whether to share a word filter. 
However, it may be the case that even a publicly available word filter still has utility as only few bad actors are motivated enough to seek to bypass a filter. This is our rationale for including built-in word filters gathered from public resources.

Similarly, participants expressed an interest in collaboratively co-creating word filters (Sec. \ref{sec:needs_share}), but poor configuration changes by one creator could disrupt the moderation for all collaborators. In addition, right now our tool only supports the cloning of existing word filters, as opposed to a subscription or forking model, where changes to the original filter can flow to subscribers.
Also, we note that many creators have a small team of volunteer or paid moderators who help them with reviewing comments; our tool at the moment only has a login for the creator of the channel. In the future, collective governance mechanisms will become increasingly important as our tool moves from being primarily used individually to author one's own filters towards more collective ways of moderating together or for each other.

\subsection{Role of Different Stakeholders}
We found in our needs analysis that much of the burden of content moderation rests on just one stakeholder---the content creator.
This is especially challenging for creators who belong to minoritized communities and receive inordinately high volumes of inappropriate comments. Previous research has examined how viewers can play a crucial role in supporting such creators by volunteering to serve as channel moderators, helping creators configure appropriate word filters, and endorsing through donations~\cite{Mahar2018squadbox}.
As demonstrated by our tool, creators themselves can also help one another by co-constructing shared resources including word filters, starting a dialogue on moderation strategies, and engaging in collective action to demand better working conditions \cite{niebler2020youtubers,chen2019youtubers}.
Drawing on our design exploration with FilterBuddy, we reflect here on how other key stakeholders, including platforms, third-party organizations and advocacy groups, and policymakers can alleviate some of the pressures and assist creators in flourishing healthy online spaces.

\subsubsection{Platforms}
From our examination of existing word filter tools on major platforms, we found that they lack many basic usability features for organization, visualization, and interactivity, and these omissions lead to frustration for content creators. 
Our subsequent development and evaluation of FilterBuddy confirms that these features are both readily implementable using standard techniques and highly desirable to creators. The question then remains why major platforms, who have considerably more resources than we do to design and develop such features, have not done so yet. 
We call upon platforms to dedicate more resources towards building more powerful creator-led tools for moderation. Alternatively, or even better, additionally, platforms could go beyond providing API access to their data and dedicate technical and human resources to foster an ecosystem of third-party tools such as FilterBuddy.
They can demonstrate their commitment to improving creators' work conditions by collaborating with academic researchers and minority support groups to build long-needed moderation tools \cite{mallari2021understanding}.


\subsubsection{Third-party organizations and advocacy groups}
As the response to FilterBuddy shows, researchers, volunteer users, and third-party organizations can play a critical role in supporting creator communities by devising and deploying novel moderation solutions. 
These groups can engage closely with marginalized content creators, understand their evolving moderation and information needs, surface them to platforms, policymakers, end-users with computational expertise and the research community, and contribute resources to address those needs.

Our participants highlighted the ability to import built-in platform categories as their favorite feature of FilterBuddy. This emphasizes the importance of offering technical resources---such as carefully curated lexicon---in addressing online hate. 
Recent work has shown how widely-adopted word filter lists such as the List of Dirty, Naughty, Obscene and Otherwise Bad Words (LDNOOBW) can harm marginalized groups, such as by censoring terms related to LGBTQ topics, due to the lack of input from members of those groups~\cite{dodge2021documenting}.
There is an opportunity here to involve minority support groups and use their domain expertise and influence to curate and publicize appropriate lexicons. 
By creating such lists and making them easy to adopt and fork in FilterBuddy, advocacy groups could 
promote social norms that discourage the use of toxic language such as racist slurs and homophobic swear words.


\subsubsection{Policymakers}
While prior HCI policy research has largely focused on the forms of public law that sweepingly regulate technology design, it is important to recognize that formal law intersects with procedures set by private firms and individual regulators in complex and nuanced ways \cite{jackson2014policy,centivany2016popcorn,kaye2019speech}. 
For example, formal law on what is considered as hate speech would likely shape the word filter categories adopted by FilterBuddy users.
It is, therefore, vital that policymakers attend to how public policy shapes technical design and social practices around moderation tools for creators.

One crucial way to empower creators is for policymakers to mandate that larger platforms must provide creators with moderation mechanisms that let them efficiently manage their channel's comments.
Such mechanisms may include the ability to author word filters, configure spelling variants, and examine their performance over time.
It has been well recognized that legislative focus on certain protected characteristics such as race, gender and religion has inadvertently excluded other vulnerable groups such as immigrants \cite{ullmann2020quarantining}; policies to mandate authoring of personalized filters can help protect such groups.
Policymakers can also incentivize platforms to enable APIs, plugins, or protocols to create a third-party marketplace of governance services~\cite{masnick_2019,wolfram_2019,2021-modpol,esber2021meta}, that will further empower content creators.

\subsection{Ethical Considerations}
Power of moderating content online can be exploited to censor important conversations \cite{gillespie2018custodians,myers2018censored,suzor2019we}. Rule-based moderation tools like FilterBuddy that allow auto-removal of comments containing certain keywords can further facilitate this process.
Deploying such tools to reduce harm, therefore, raise enduring, familiar tensions between libertarian tendencies (e.g., freedom of speech) and authoritarian practices (e.g., censorship) \cite{ullmann2020quarantining,dworkin2009extreme,waldron2012harm,waldron2017conditions,weinstein2017hate}.
However, we note that the ability to block configured keywords already exists on most platforms, including YouTube, in the form of word filters. In fact, FilterBuddy offers many analytic and visual aids to avoid setting up configurations that result in false positives, so that well-meaning creators do not inadvertently censor valuable comments.
Yet, third-party moderation tools (e.g., Twitter blocklists \cite{jhaver2018blocklists,geiger2016}) have previously been exploited in ways unanticipated by their creators.
Therefore, we are committed to conducting a regular oversight of who uses FilterBuddy and how it is being used.

\section{Limitations and Future Work}

Going forward, we plan to implement additional features that study participants requested in our evaluation (Section \ref{sec:evaluation}). This includes 
incorporating additional types of spelling variants,
creating a time selector and video selector for graphs, and configuring more complex word filters that identify, for example, the co-occurrence of multiple phrases in the same comment.
After additional development to improve security, we will release FilterBuddy as a public hosted tool and open-source its code to encourage further development. 

Currently, FilterBuddy is a YouTube-only tool; however, it could be extended to other platforms that support 
creator-oriented communities with little additional development. The YouTube API features that we use, including
comment retrieval and comment removal, have uneven but growing support across other major platforms. 
Many of FilterBuddy's features are also valuable and much needed in other moderation contexts. For example, prior research has shown that volunteer moderation teams on sites like Reddit often struggle with using moderation tools that are difficult to configure and have lack of visibility into the operations of such tools \cite{jhaver2019automated}; FilterBuddy offers to address these needs.

FilterBuddy relies on analyzing sequences of written words to automatically detect undesirable content. This entirely text-based approach is insufficient to regulate online communication that is frequently and increasingly multimodal. Thus, it is important to incorporate and build upon prior research on multimodal approaches to hate speech detection in the future \cite{hosseinmardi2015detection,zhong2016content}.
Further, since FilterBuddy is a regular expression-based tool, it is limited in its ability to detect posts that use more subtle or complex linguistic strategies to propagate hate speech. Similarly, it might have limited utility in addressing some forms of undesirable content such as political or health misinformation.
Yet, we expect that this tool will introduce friction \cite{chandrasekharan2020quarantined} to successfully posting inappropriate posts and act as a powerful deterrent against bad actors, thereby delimiting their submission of such posts.
Prior research has shown that poorly implemented blocklists can disproportionately remove text from and about minority individuals and exacerbate existing inequalities \cite{dodge2021documenting,rosa2019looking,pinsof2017effect}. We hope that FilterBuddy's analytic and visualization features to configure more accurate word filters would help minimize such harms.

Since we evaluated FilterBuddy using a small user study, we do not yet have empirical evidence about how such a system would work in the wild. 
While our user study helped us understand how creators would use FilterBuddy and suggested promising design directions, an in-the-wild study would clarify new aspects, such as determining possible incentives towards sharing word filters with other creators, using multiple categories for the same content type, and defending against bad-faith actors.
We suspect that our user study may have introduced some biases, e.g., participants may have overstated their desire for control, which may alter as the creators use our tool in the wild and get overworked; we are excited to see if this calculus changes in a long-term field deployment.
Even in our small user study, participants devised a number of use cases for creating FilterBuddy categories that we had not anticipated. Therefore, an in-the-wild study would also help us identify further utility of this tool.
We are also excited to study the sharing practices among YouTube creators and further strengthen the security and utility of sharing mechanisms FilterBuddy provides. 


\section{Conclusion}
In this work, we study the current practices of how online content creators use rule-based moderation tools, finding from 19 interviews that many creators find existing tools cumbersome to use, consider it difficult to come up with keywords to configure, and want analytic information that can guide their configuration and review of these tools. We developed FilterBuddy, a word filter tool that helps content creators better address their moderation needs by offering new visualizations, configuration assistance, and ready-made categories.
As opposed to the reactive, too-little-too-late approach primarily used by platforms to decrease online harm, FilterBuddy intercepts potentially problematic posts at an earlier stage of the process, before they have been read by the content creators or their audiences. 
From a user study, we found that creators appreciate the features that FilterBuddy provides, desire to share their configurations with other creators, and show resistance to replacing the rule-based configurations of this tool with solely ML-based approaches.

\begin{acks}
Omitted for review.
\end{acks}

\bibliographystyle{ACM-Reference-Format}
\bibliography{moderation_references,00-references}

\end{document}